# Exceptional Point Engineered Glass Slide for Microscopic Thermal Mapping

Han Zhao[1], Zhaowei Chen[2], Ruogang Zhao[2]* and Liang Feng[3]*

[1]*Department of Electrical and Systems Engineering, University of Pennsylvania, Philadelphia, PA 19104, USA*

[2]*Department of Biomedical Engineering, The State University of New York at Buffalo, Buffalo, NY 14260, USA*

[3]*Department of Materials Science and Engineering, University of Pennsylvania, Philadelphia, PA 19104, USA*

*Emails: fenglia@seas.upenn.edu; rgzhao@buffalo.edu

Thermal sensing with fine spatial resolution is important to the study of many scientific areas. While modern microscopy systems allow easy optical detection at high spatial resolution, their intrinsic functions are mainly focused on imaging but limited in detecting other physical parameters, for example, mapping thermal variations. Here, with a coating of an optical exceptional point structure on a conventional glass slide, we demonstrate a novel low-cost and efficient multifunctional microscope slide, supporting real-time monitoring and mapping of temperature distribution and heat transport in addition to conventional microscopic imaging. The square-root dependency associated with an exceptional point leads to enhanced thermal sensitivity that is critical for the precise measurement of the temperature. With a microscale resolution, real-time thermal mapping is conducted, showing dynamic temperature variation in a spatially-defined area. Our strategy of integrating low-cost and efficient optical sensing technologies on a conventional glass slide is expected to enable simultaneous detection of multiple environmental parameters, thereby leading to new opportunities for improved experimental control in many scientific research disciplines.

Optical microscopes produce magnified images of all types of specimens, enabling direct observation of microscale details of the objects. It has become a fundamental tool advancing the modern photonic and biological sciences. Beyond the conventional optical imaging, real-time inspection and mapping of various physical parameters are highly desired with microscale spatial resolution [1-4], including temperature, humidity, pH, particle impurity and atmospheric pressure. However, despite improved spatial resolution, the state-of-the-art microscope systems are still intentionally designed for tomography-type imaging, rather than simultaneously mapping other important physical parameters [5-9]. Among many environmental parameters crucial to chemical and biological measurements is the temperature [4]. It remains a grand challenge to measure the temperature distribution of a sample with high sensitivity and high spatial-resolution. For example, the spatial resolution of the conventional thermo-couplers is typically on the order of millimeters [10], which is too coarse for microscopic temperature detection. The reliabilities of terahertz or infrared thermal mapping techniques are, unfortunately, significantly affected by the reflection from the specimen surfaces as it is difficult to be distinguished from the optical signal radiated from the heat source [11-14]. Although plasmonic resonant structures were attempted to characterize the thermal perturbation where the sensitivity is enhanced by plasmonic resonance-induced strong field confinement, the raw thermal sensitivity is limited by the linear proportion to the perturbation strength [15-17]. Additionally, the required wavelength scanning is time-consuming and thus significantly prevents its practical applications for efficient real-time monitoring and mapping of the ambient thermal distribution of the target specimen.

Recently, the emergence of non-Hermitian optics has revolutionized a strategic control of light transport by exploring the complex dielectric permittivity on its entirety [18-22]. Non-Hermitian singularities, known as exceptional points (EP), can arise, featured by simultaneous coalescence of both eigenvalues and eigenstates in the complex eigen spectrum. EPs are different from the energy-degenerate diabolic points (DPs) in conventional optical resonant structures that possess only the degenerate eigenvalues. Subjected to a weak parametric perturbation, the EP degeneracy can be lifted, resulting in unique eigenvalue splitting proportional to the square-root of the perturbation strength. When exploited as a weak transduction signal in sensor applications, such square-root energy splitting represents unprecedented enhancement of raw sensitivity superior to the linear response from the DP degeneracy [23-26].

Here, without modifying the microscope system, we demonstrate a low-cost thermal sensitive microscope slide for highly-distributed temperature mapping and real-time monitoring of heat transport, in addition to the conventional optical imaging. By utilizing the enhanced sensitivity associated with non-Hermitian EPs, a multilayer EP structure was designed and coated on a conventional microscope slide to sense the temperature variation. Under the ambient thermal perturbation, the polymer layer in the EP structure deforms due to the transferred heat, causing the lifting of EP degeneracy and consequently the increase of reflection under a square-root relation. Through the reflection measurements at the initial EP wavelength, the temperature distribution on the glass slide can be gauged with high spatial resolution, facilitating efficient thermography in addition to conventional tomographic imaging in an intact microscope system.

The most important feature of the multifunctional slide is its compatibility and easy implementation with a microscope. As shown in Fig. 1a, a thermal-sensitive glass slide is conveniently mounted on a conventional epi-fluorescent microscope system. While maintaining the transmitted-light imaging mode intact, a monochromatic laser light is applied to map the reflection variation at the EP wavelength. As the most cost-efficient and widely used laser source in many routine laboratory settings, a He-Ne laser at the wavelength of 632.8 nm, which can be seamlessly integrated with any microscope, is chosen in our work to probe the optical transduction for thermal imaging.

Upon normal incidences of the He-Ne probe laser on a two-port optical system (note that a microscope slide is a typical two-port system), its optical characteristics can be described using the scattering matrix

$$S = \begin{pmatrix} t & r_b \\ r_f & t \end{pmatrix}, \quad (1)$$

where $r_b$ and $r_f$ are the reflection coefficients in backward and forward directions, respectively, and $t$ denotes the transmission coefficient that is the same in both directions due to reciprocity. The scattering matrix can be derived from the standard optical transfer matrix method [27]. The eigenvalues of the scattering matrix are $\upsilon_{1,2} = t \pm \sqrt{r_f r_b}$. To achieve the EP where two eigenvalues become degenerate, it is clear that the eigenvalue splitting has to reach zero, i.e., $2\sqrt{r_f r_b} = 0$. Such coalescence of the scattering eigenstates corresponds to the previously demonstrated unidirectional reflectionless condition, $r_b \neq r_f = 0$ [22,28,29].

To demonstrate a glass slide of such a unidirectional reflectionless EP response, we conduct a 3-layer hetereostructure design with a delicate interplay of a thermally deformable polymer layer of polymethyl methacrylate (PMMA) sandwiched between two Au films (Fig. 1b). By optimizing the respective thickness of each layer, the EP-supported unidirectional reflectionless light transport is obtained for the probe laser, i.e. at the wavelength of 632.8 nm. At the initial state of room temperature, the EP condition results in complete reflection darkness in the forward direction due to the coalescence of the scattering eigenstates. Under temperature perturbation, the local thickness of polymer varies due to thermal expansion/suppression, which breaks the EP condition, lifting the EP degeneracy and thus causing a drastic reflection enhancement. Therefore, spatially imaging the reflection change is equivalent to resolving and mapping the thermal response of the polymer layer, which can be further converted to a microscopic image of temperature distribution.

It is important to note that despite the coating of the multilayer EP structure, the glass slide is transparent under transmitted light, evidently indicated by the visible landscape behind the slide (Fig. 1c). Hence, in addition to the new temperature detection function, the glass slide still supports the conventional imaging function of a microscope under the transmitted light mode.

Another critical property of the multifunctional slide is its thermal sensitivity. To precisely map the temperature distribution and variation, it is important to demonstrate high thermal sensitivity. In our designed multilayer EP structure, sensitivity enhancement arises from the intrinsic square-root dependency of the scattering eigenvalues of the system as the temperature perturbation drives the system across the EP [25,26].

The EP and its associated phase transition are validated through the calculated amplitude spectrum of the generalized reflecton coefficiet that is half the scattering eigenvalue splitting: $\Delta\upsilon/2 = \sqrt{|r_f r_b|}$ (Fig. 2a). Around the EP, the splitting of the squre-root scattering eigenvalue sharpens the phase transition in wavelengths across the EP, which is fundamentally distinguished from the linear variation in its counterpart of Hermitian degeneracies (i.e. DPs). Such an abrupt phase transition at the EP can also be clearly revealed if measuring only the forward reflection coefficient $|r_f|$, where a quasi-linear curve is observed around the EP (Fig. 2a). This observation helps significantly simplify our experiments. Instead of analyzing the eigenvalue splitting that requires reflection measurements from both directions, only forward reflectoin needs to be characterized to assess all the EP-related properties (Fig. 2b).

To establish a real-time optical transduction mechnism in thermal sensing, the correlation of the ambient temperature and the scattering eigenvalue splitting is measured as a function of the amplitude of forward reflection at the fixed EP wavelength. Enabled by performing the perturbation analysis, assume the transferred heat causes the thickness variation of the polymer (PMMA) layer $L$ by a sufficiently small deformation $\Delta L$. The scattering matrix subjected to the thermal-induced perturbation becomes

$$S = S_0^{EP} + \Delta L \begin{pmatrix} \frac{\partial t}{\partial L} & \frac{\partial r_b}{\partial L} \\ \frac{\partial r_f}{\partial L} & \frac{\partial t}{\partial L} \end{pmatrix} = \begin{pmatrix} t_0^{EP} + \Delta L \frac{\partial t}{\partial L} & r_b^{EP} + \Delta L \frac{\partial r_b}{\partial L} \\ r_f^{EP} + \Delta L \frac{\partial r_f}{\partial L} & t_0^{EP} + \Delta L \frac{\partial t}{\partial L} \end{pmatrix}, \quad (2)$$

where $r_b^{EP} \gg r_f^{EP} = 0$, referring to the initial EP unidirectional reflectionless condition. Thermal deformation lifts the EP degeneracy with scattering eigenvalue splitting of

$$\Delta \upsilon = 2\sqrt{\Delta L \frac{\partial r_f}{\partial L}\left(r_b^{EP} + \Delta L \frac{\partial r_b}{\partial L}\right)}. \quad (3)$$

Because of the large $|r_b^{EP}|$, the square-root relation is revealed as $\Delta \upsilon^{EP} = 2\sqrt{\frac{\partial r_f}{\partial L} r_b^{EP}} \cdot \sqrt{\Delta L}$, where the sensitivity can be fully characterized by only the derivative of forward reflection. Note that for a DP degeneracy, Eq. (3) evolves to a linear relation with the thickness perturbation of $\Delta \upsilon^{DP} = 2\Delta L \frac{\partial r_f}{\partial L}$ since the Hermitian condition regulates simultaneous reflection vanishing in both dierections (i.e. $r_b^{DP} = r_f^{DP} = 0$). Such direct contrast of eigenvalue splitting between the designed EP structure and a typical DP structure (such as a PMMA anti-reflection film) reads

$$\frac{\Delta\upsilon^{\text{EP}}}{\Delta\upsilon^{\text{DP}}} = \sqrt{\frac{r_b^{\text{EP}}}{\Delta L \cdot \partial r_f / \partial L}}. \quad (4)$$

Under weak perturbation, the relation $r_b^{\text{EP}} \gg \Delta L \cdot \partial r_f / \partial L$ ensures significant sensitivity improvement. Specifically, starting from the degeneracy points, the unit temperature change of 1 K leads to EP splitting of $\Delta\upsilon^{\text{EP}} = 0.166$, offering orders of magnitude enhancement compared with the merely detectable DP splitting of $\Delta\upsilon^{\text{DP}} = 0.004$ under the same thermal perturbation (Fig. 2c). Fig. 2d shows the measured and calibrated relationship of the temperature variation with the refelction signal from a He-Ne probe laser. The observed enhanced thermal sensitivity of 0.0173 K$^{-1}$ can enable precise readings of temperature on the glass slide.

    The purpose of the thermal mapping is to retrieve detailed temperature information of the specimen, along with conventional microscopic imaging, to enable the analysis of the correlation between multiple physical parameters. To meet this objective, the desired technique must support thermal mapping, operated in a highly-distributed manner with a high spatial resolution comparable to the resolution of the microscope. The ultimate spatial resolution of thermal mapping is only limited by the thermal diffusion induced spatial overlapping of temperature distribution [30], and is evaluated to be at a 10-μm scale in our device. Here, we have validated such a highly-distributed thermal mapping function of our multifunctional microscope slide and its associated microscale spatial resolution. Spatially distributed local thermal sources were effectively generated by optically casting and focusing a 3×3 square-latticed hole array onto the central layer of PMMA to locally heat up and expand the polymer through the backside of the slide. To avoid the interference with the visible measurements, we used a 10-nanosecond pulsed laser with a center wavelength of 1064 nm. Through demagnification (see Supplementary Information for the details of the experimental setup), 9 microscale laser spots were created with a spot diameter averaged at 30 μm, as shown by the corresponding transmission image at the wavelength of 1064 nm (Fig. 3a). While only a small fraction of laser light can be absorbed by the polymer, the heat accumulated from absorption still induces temperature increase

around those 9 spots, which locally enlarges the thickness of the central polymer layer. Due to thermal diffusion in the EP multilayer structure, the absorbed heat from laser spots spreads and raises the temperature evanescently within the vicinity of approximately 4 μm. As a result, only the local condition around the laser spots deviates away from the EP operation designed for room temperature, causing the reflection variation of the probe light from the He-Ne laser by which the temperature change in the vicinity of the laser spots can be mapped. To completely eliminate the adverse effect on thermal mapping due to the transmission of the casted heating laser beam, a bandpass filter at 500 nm – 700 nm was placed in front of the camera to collect and image only the reflected probe light of the He-Ne laser beam to precisely read the thermal distribution on the microscope slide. It is worth noting that the accumulated heat as well as its caused variation of the local temperature and deformation of the polymer thickness are all linearly dependent of the power of the casted pulsed laser. Therefore, by varying the power of the pulsed laser, we observed a linear growth in forward reflection of the probe light, which was further converted to a function of temperature through the calibrated temperature-reflection correlation (Fig. 3b). The thermal maps retrieved from the probe light are displayed in Fig. 3c with different averaged incident heating laser power. Remarkably, the imaged thermal distributions spatially resolve the pattern of the casted heating pulsed laser spot with feature size at 30 μm.

Another technical advantage of the multifunctional microscope slide is its high temporal-resolution for real-time monitoring of temperature distribution and evolution in samples. As many temperature-critical processes occur under a solvent environment [31-33], we mimicked such an environment by injecting hot water on the glass slide, and experimentally show the dynamic evolution of the water temperature during heat diffusion into the ambience.

To confine hot water in a local area for imaging, a water reservoir was built on the slide using a block of polydimethylsiloxane (PDMS) with a hollow octagonal cavity in the middle, which was then bonded on top of the layered EP structure, as schematically depicted in Fig. 4a. As our aforementioned discussion, while the thermal mapping is performed in a reflection mode mapping the temporal evolution of the temperature distribution of injected water, the slide is still transparent to support microscopic imaging using white light in a transmission mode, where the boundary of PDMS divides the entire field into two regions: an area in the reservoir where hot water injection induces the temperature change and the other area of stabilized temperature covered by PDMS (Fig. 4b). It should be noted since the top cladding changes from air to water after the hot water injection, the original EP design becomes invalid such that we reconfigure the multilayer structure to support zero-forward reflection for the probe light to function in the EP condition.

As a control experiment, room temperature water was first injected into the reservoir. Due to their similar refractive indices, water and PDMS provide almost equivalent cladding environments. In this case, the temperature of the glass slide surface maintains at room temperature during the water injection. The detected reflection in the field of view stays uniform at its resonance minimum of the designed EP condition in both water and PDMS regions (Fig. 4c).

To observe temporal evolution of the heat transfer process, we injected hot water at approximately 43 °C and real-time monitored and mapped the temperature distribution. Figs. 4d-4g show the measured instantaneous thermal maps at different time delays. Right after injecting a droplet of hot water, because of the high thermal conductivity of the thin gold layer, the underneath PMMA layer in the water reservoir region immediately responds with a thermal expansion. As a consequence, the detected reflection significantly increases from the EP resonance minimum, indicating the corresponding local temperature change. Nevertheless, only a negligible amount of heat from hot water can horizontally transfer through the PDMS wall of a much lower thermal conductivity, barely varying the thickness of the PMMA layer under PDMS. This leads to an evanescent decay of thermal perturbation across the edge of the reservoir, creating a spatial temperature gradient. As the heat dissipates in the fluxing direction from the water reservoir to the PDMS cladding until the thermal equilibrium condition is reached, the temperature of water keeps dropping and so is the temperature of the PMMA layer, leading to the contraction of the PMMA layer back to the initial EP condition and its resulted decrease in reflection back to the initial resonance EP minimum. This process has been dynamically recorded in which the spatial temperature gradient gradually retracts in time towards the center of the reservoir. Here, the designed multifunctional microscope slide successfully enables real-time monitoring of the dynamic evolution of a heat transfer process, promising *in-situ* control of temperature in the applications where temperature monitoring is necessary. For example, it may play an important role in protein biosynthesis where enzyme activities are considerably affected by heat generation [34,35].

In summary, utilizing the enhanced sensitivity arising from an optical EP, we have demonstrated a novel multifunctional glass slide facilitating simultaneous microscopic mapping and monitoring of temperature distribution and thermal variation. The results evidently show that the microscope slide functions as an efficient temperature monitor with enhanced thermal sensitivity and micro-scale spatial resolution. The described optical transduction mechanism can revolutionize the functionality of ubiquitous glass slides beyond their conventional mechanical supporting of specimen, promising a low-cost microscope-integrated sensing technology for improved controllability on multi-laboratory environmental parameters.

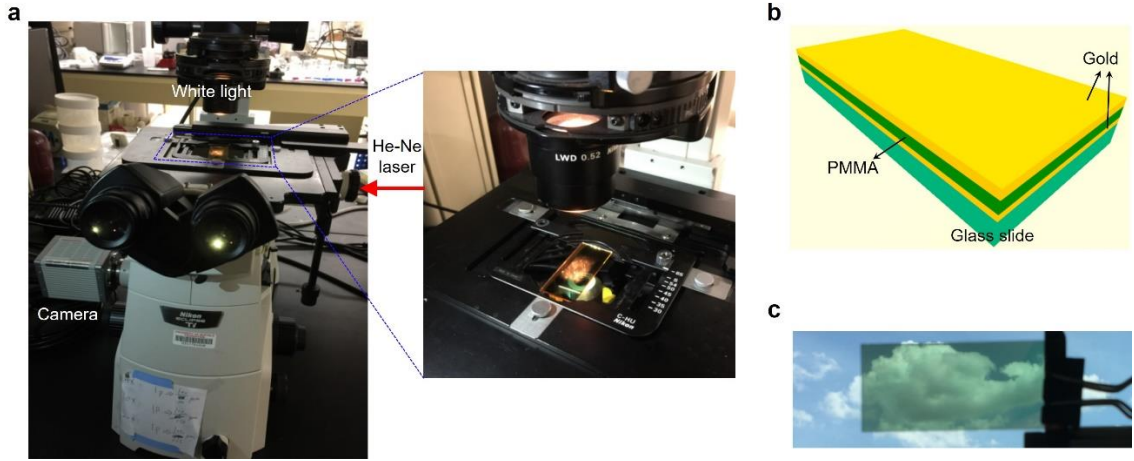

FIG. 1. Thermal sensitive microscope slide engineered at exceptional point: (a) A microscope system with the devised thermal sensitive multifunctional glass slide. While the white light source is for normal microscopic images, the He-Ne laser (labeled by red arrow) is used as the incidence for the thermal mapping. Inset: zoom-in of the glass slide in the micrsope system. (b) Schematic drawing of the thermal sensitive glass slide engineered at optical exceptional point. A gold-PMMA-gold three-layer structure is deposited on silica glass slide. (c) Transparency of the multifunctional glass slide. Revealed by the observable landscape behind the glass slide, the transmission to white light is high enough for optical microscope characterization, while thermal sensing is added as additional function to the glass slide.

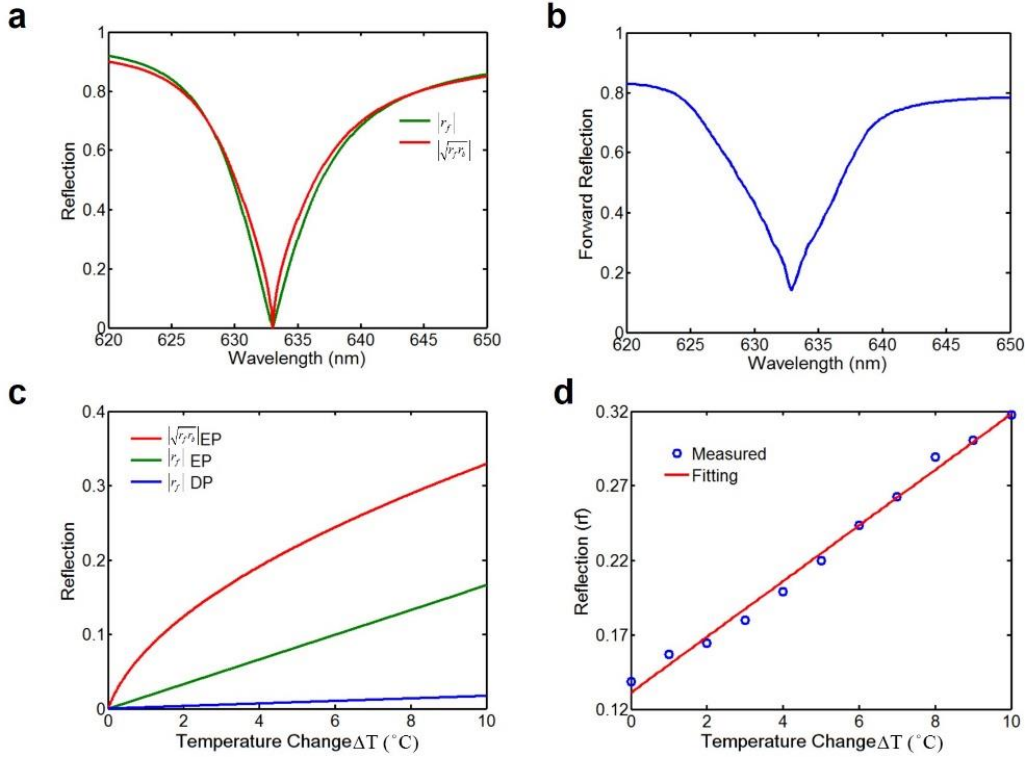

FIG. 2. Optical characterization of enhanced thermal sensitivity enabled by exceptional point: (a) Calculated spectra of half the scattering eigenvalue splitting ($\Delta\upsilon/2 = \sqrt{|r_f r_b|}$, red curve) and forward reflection ($|r_f|$, green curve). The EP degeneracy featured by zero scattering eigenvalue splitting is designed at 632.8 nm by optimizing the respective thicknesses of Au and PMMA layers. The abrupt EP phase transition can be fully characterized by the quasi-linear response of forward reflection with negligible difference. (b) Experiemtally measured spectrum of forward reflection, where the minimum forward reflection at the exceptional point wavelength is measured at 0.14 (corresponding to a reflectance of approximely 2% only) due to imperfection in fabrication. However, the sharp EP phase transition is still observed. (c) Theoretical temperature responses of scattering eigenvalue splitting (red curve) and forward reflection (green curve) at 632.8 nm, compared to the forward reflection of the pure PMMA anti-reflection DP degeneracy (blue curve). Near the EP degeneracy, the scattering eigenvalue splitting evolves in square-root dependency on the temperature incremental, providing orders of enhancement in sensitivity. Characterized by the derivative of forward reflection, the raw thermal sensitivity in our designed EP structure averages 0.018 K$^{-1}$, in stark contrast to the low

sensitivity of 0.00196 K$^{-1}$ from DP degeneracy. (d) Calibration of temperature-reflection linear response. Although the initial state deviates from an ideal EP condition (zero reflection), the linear response is measured with a thermal sensitivity of 0.0173 K$^{-1}$, which is consistent with the theoretical prediction of 0.018 K$^{-1}$ in (d).

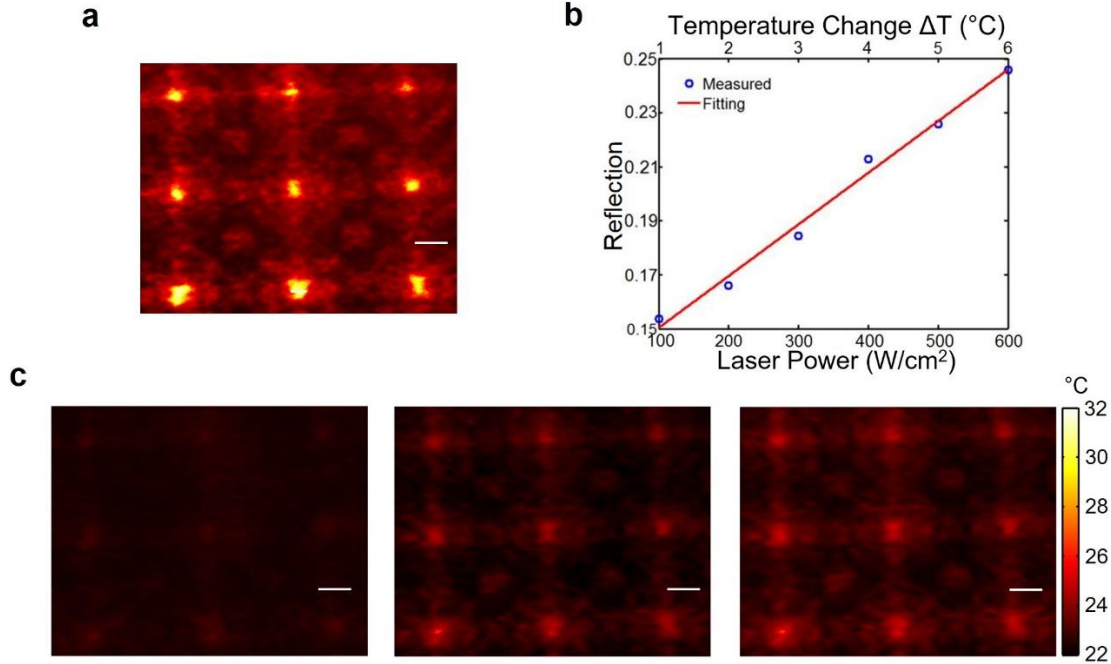

FIG. 3. Thermal mapping of spatially resolved pulse laser heating source: (a) Spatially distributed thermal source of 3×3 spot array revealed by transimission of the incident 1064 nm pulse laser on the glass slide. The laser spots with diatmeter averaged at 30 μm are created by imaging square-latticed hole array on the central PMMA layer through a demagnifying optical setup. As the glass slide absorbs incident pulse laser power, the transferred heat thermally diffuses in the PMMA layer and locally raises the temperature, which results in expansion of PMMA thickness and lifting of EP degeneracy. (b) The correlation of the pulse laser power density and forward reflection of the probe He-Ne laser beam measured at the center spot. With increasing incident power of the heating source, forward reflection of probe He-Ne laser grows linearly from the minimum at the EP condition. (c) Spatially resolved thermal mappings of the heating laser spot array revealed by forward reflection of He-Ne laser beam. Panels from left to right were obtained at increasing incident pulse laser power density at 100 W/cm$^2$, 300 W/cm$^2$ and 500 W/cm$^2$, respectively. Scale bar in (a) and (c), 50 μm.

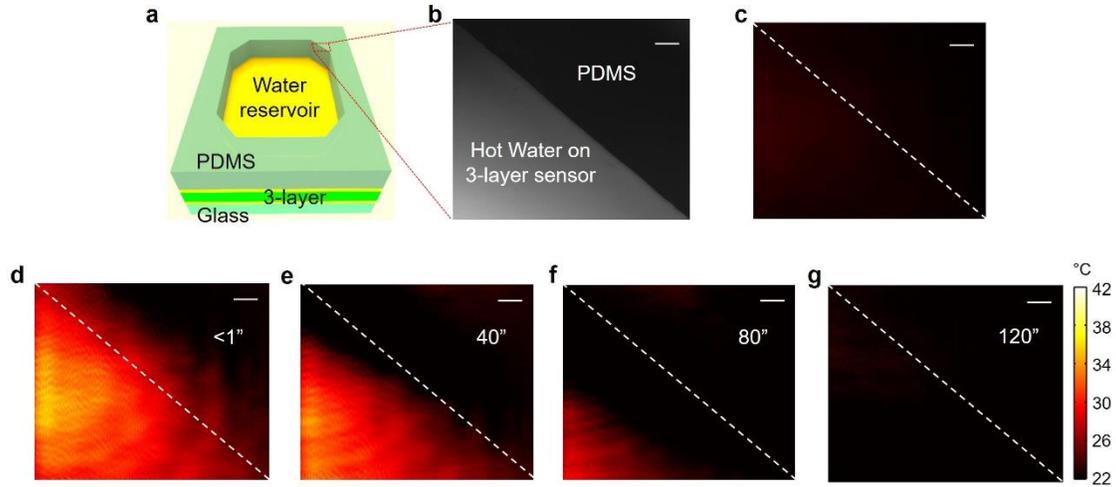

FIG. 4. Transient thermal mapping of hot water injection: (a) Schematic drawing of the thermal sensitive microscope slide with bonded PDMS octagonal water reservoir for hot water injection. (b) Microscope image of the top right corner of the water reservoir by transmitted light source. The boundary divides the water and PDMS regions, demonstrating transmitted light imaging capacity of the multifunctional glass slide. (c) Thermal mapping of the reservoir edge area when the reservoir is filled with room temperature water as a control experiment. In the absence of heat transfer, the uniform temeperature distribution from detected reflection reveals the initial EP condition in both water and PDMS regions. (d)-(g), Transient thermal mappings at different time points after 43 °C hot water injection using forward reflection of the He-Ne laser beam. The instanteneous jump of He-Ne laser reflection in the resorvior and the evanescent tail towards the PDMS region evidently show the in-plane heat transfer process. As the heat diffuses, the detected reflection damped gradually anti-parellel to the horizontal heat fluxing direction from water to PDMS until the uniform map at thermal equilibrium reached 2 minutes after water injection, manefesting the evolution of spatial temeperature distribution in real time. Dashed lines in (c)-(g) represent the boundary of water and PDMS. Scale bar in (b)-(g), 100 μm.